\documentclass{iopart}

\usepackage{epsfig}
\usepackage{amscd}
\usepackage{graphicx,xspace}
\input{epsf}

\include{latexcommands}

\def\be{\begin{equation}}
\def\ee{\end{equation}}

\def\bea{\begin{eqnarray}}
\def\eea{\end{eqnarray}}

\def\e{\epsilon}

\def\k{\kappa}
\def\a{\alpha}

\def\g{\gamma}
\def\ct{{\tilde c}}

\begin{document}

\title[One-particle  density matrix of 1D anyons]{One-particle  
density matrix and momentum
distribution  function of one-dimensional anyon gases}

\author{Raoul Santachiara${}^{1}$ and Pasquale Calabrese${}^{2}$}

\address{$^{1}$ CNRS-Laboratoire de Physique Theorique de l'Ecole Normale 
Superieure, 24 rue Lhomond, 75231 Paris, France.}

\address{$^{2}$Dipartimento di Fisica dell'Universit\`a di Pisa and INFN, 
Pisa, Italy}

\begin{abstract}
We present a systematic study of the Green functions of a one-dimensional
gas of impenetrable anyons. We show that the one-particle density matrix is the
determinant of a Toeplitz matrix whose large $N$ asymptotic is given by the
Fisher-Hartwig conjecture. 
We provide a careful numerical analysis of this determinant
for general values of the anyonic parameter, showing in full details the
crossover between bosons and fermions and the reorganization of the
singularities of the momentum distribution function. 

We show that the one-particle density matrix satisfies a Painlev\'e $VI$
differential equation, that is then used to derive the small distance and large
momentum expansions. We find that the first non-vanishing term in this
expansion is always $k^{-4}$, that is proved to be true for all couplings in 
the Lieb-Liniger anyonic gas and that can be traced back to the presence of 
a delta function interaction in the Hamiltonian.

\end{abstract}

\maketitle

\section{Introduction}

Generalized anyonic statistics, which interpolate continously between bosons 
and fermions, are considered one of the most remarkable breakthrough of 
modern physics.
In fact, while in three dimensions particles can be only bosons or fermions, 
in lower dimensionality they can experience exchange properties intermediate 
between the two standard ones \cite{anyons}. 
In two spatial dimensions, it is well known that fractional braiding
statistics describe the elementary excitations in quantum Hall effect, 
motivating a large effort towards their complete understanding. 
Conversely the study of one-dimensional (1D) anyons is still at an embryonic 
stage, although recent interesting proposals for topological quantum 
computations based also on 1D anyons \cite{tqc}.

In one dimension, anyonic statistics are described in terms of fields that 
at different points ($x_1 \neq x_2$) satisfy the commutation relations   
\bea
\Psi_A^\dag(x_1)\Psi_A^\dag(x_2)&=&e^{i\k \pi\e(x_1-x_2)} 
\Psi_A^\dag(x_2)\Psi_A^\dag(x_1)\label{ancomm}\,,\\
\Psi_A(x_1)\Psi_A^\dag(x_2)&=&e^{-i\k \pi\e(x_1-x_2)} 
\Psi_A^\dag(x_2)\Psi_A(x_1)\,,
\nonumber
\eea
where $\e(z)=-\e(-z)=1$ for $z>0$ and $\e(0)=0$. $\k$ is called statistical
parameter and equals $0$ for bosons and $1$ for fermions.
Other values of $\k$ give rise to general anyonic statistics ``interpolating''
between the two familiar ones.

A few 1D anyonic models have been introduced and investigated 
\cite{k-99,bgo-06,bg-06,g-06,ssc-07,zw-92,bg-06b,cm-07,pka-07,an-07,lm-99,it-99,kl-05,fibo,g-07,zw-07,o-07}. 
In this paper we consider the anyonic generalization of the Lieb-Liniger gas 
defined by the (second-quantized) Hamiltonian
\be
\fl H=\frac{\hbar^2}{2M}\int_0^L dx \partial_x\Psi_A^\dag(x) \partial_x\Psi_A(x)
+c \int_0^L dx \Psi_A^\dag(x) \Psi_A^\dag(x) \Psi_A(x) \Psi_A(x)\,, 
\ee
which describes $N$ anyons of mass $M$ on a ring of length $L$ interacting 
through a local pairwise interaction of strength $c$ (in what follow, we fix
$2M=\hbar=1$). 
In first-quantization language, the Hamiltonian is
\begin{equation}
H=-\sum_{i}^{N}\frac{\partial^2}{\partial x_{i}^{2}}+
2c\sum_{1\leq i<j\leq N}\delta(x_i-x_j),
\label{ALL}
\end{equation}
where now is the $N$-anyons wave-function 
$\Psi^\k(x_1,x_2,..,x_N)$ to exhibit a generalized symmetry
under the exchange of particles
\begin{equation}
\Psi^\k(\cdots x_j,x_{j+1}\cdots)=
  e^{i \pi \k \e(x_{j+1}-x_j)} \Psi^\k(\cdots x_{j+1},x_j\cdots).
\label{anyon_exhange}
\end{equation}
For $\k = 0$ the model reduces to the bosonic Lieb-Liniger \cite{LL}, 
while for $\k=1$ to free fermions. 
The physics only depends on the dimensionless parameter $\g=c/\rho_0$ (with 
$\rho_0$ the mean density $\rho_0=N/L$) and not on the two parameters
separately (a part from obvious scaling factors). The interest in this model
is mainly due to the fact that it is the simplest solvable model of
interacting anyons and in fact it has been shown that it has a 
solution in terms of Bethe Ansatz \cite{k-99,bgo-06}. 
Furthermore 1D exactly solvable
models are in a renascent age after the experimental realizations of trapped
1D atomic gases in the last few years \cite{1dexp}. 
There is also a proposal to engineer an anyonic gas
by trapping bosons in a rapidly rotating trap \cite{par}. 

However, as well known for the bosonic counterpart, the Bethe Ansatz gives a
precise characterization of the spectrum of the model and of the full 
thermodynamic, but does not allow to calculate the correlation functions. 
More complicated methods employing an algebraic formalism \cite{Kbook} 
(eventually joined to numerical calculations \cite{cc-06}) must be used 
in order to extract the correlation functions at arbitrary distances.
This with one important exception: the case of impenetrable particles, i.e. 
$c=\infty$, that is obtainable with an anyon-fermion mapping \cite{g-06}.

Here we exploit this mapping to give a representation of the one-body density
matrix
\be
\rho^\k_N(x)=\langle\Psi^{\k\dag}_A(x) \Psi^\k_A(0)\rangle\,,
\ee
in terms of the determinant of a Toeplitz matrix (completing the work started
by one of us in Ref. \cite{ssc-07}) whose large $N$ asymptotic is given 
by the Fisher-Hartwig conjecture.
We present a careful numerical analysis of $\rho^\k_N(x)$ and of the Fourier 
transform (known as momentum distribution function) for general values of 
the anyonic parameter $\k$ showing in full details the crossover between
bosons and fermions, that is highly non trivial because it involves a
reorganization of the singularities of the momentum distribution function. 

Furthermore we show that $\rho^\k_N(x)$ satisfies a Painlev\'e 
VI differential equation, that is the same as for impenetrable bosons
\cite{ffgw-02} but with different boundary conditions.
This differential equation allows a straightforward derivation of the small $x$
expansion of $\rho^\k_N(x)$ that can be used to derive the large momentum
expansion of the momentum distribution function. The first non-vanishing term
in this expansion is always $k^{-4}$ a fact that is proved to be valid for all 
couplings $c$ in the Lieb-Liniger gas and that can be traced back to the 
presence of a delta function interaction in the Hamiltonian.

After the completion of this manuscript, a complementary approach to the same
problem was used by Patu, Korepin and Averin \cite{pka-08}.
They derived $\rho^\k_N(x)$ in the thermodynamic limit as a Fredholm 
determinant generalizing the Lenard result for bosons \cite{l1}.
However, despite the fact that we both consider the same correlation,
the two approaches are complementary. In fact, our method allows to study 
effectively systems with finite numbers of particles and it is particularly 
suited for asymptotic expansions. The method of Ref. \cite{pka-08}
allows instead for a direct generalization to finite
temperature that in our approach is very cumbersome. The complementarity of the
two approaches is highlighted by the fact that there is a single common
formula in the two manuscripts (namely Eq. (\ref{Kform}) 
below, derived in very 
different ways). We stress (as done in Ref. \cite{pka-08}) 
that an explicit proof of the equivalence of the two representations
of the anyonic correlation function is still an open problem, as in the case
of bosons. 

The paper is organized as follows. 
In Sec. \ref{sec2} we present the determinant form for $\rho^\k_N(x)$
and derive its asymptotic behavior for large distance by means of the
Fisher-Hartwig conjecture. In Sec. \ref{sec3} we present the numerical 
results that allow for a
characterization of the crossover from bosons to fermions. In Sec. \ref{sec4}
we prove that $\rho^\k_N(x)$ satisfies a second order differential equation,
that in the next section \ref{sec5} is used to derive the small distance and
large momentum expansions. In Sec. \ref{sec6} we show that the power-law tail
of the momentum distribution function is generally valid for the 
Lieb-Liniger model. Finally, in Sec. \ref{sec7}, we discuss critically our
results and possible future investigation.

\section{Ground-state function and one-particle density matrix}
\label{sec2}

In the limit of impenetrable anyons, i.e. $c\to\infty$,
the $N$ anyons ground-state wave function can be easily written down by
imposing that two anyons should not occupy the same position.
As shown in Ref. \cite{g-06}, the ground-state
$ \Psi^\k_0(x_1,\cdots,x_N)$ is then
\begin{equation}
 \Psi^\k_0(x_1,\cdots,x_N)=\left[\prod_{1\leq i < j
      \leq N} A(x_j-x_i) \right]\Psi^1_0(x_1,\cdots,x_N),
  \label{AF}
\end{equation}
where $\Psi^1_{0}(x_1,\cdots,x_N)$ is the ground-state function of 
$N$ free fermions. In the following we always consider the case an odd number
of particles $N$, which 
corresponds to a non-degenerate ground state. In this case we have
\bea
\Psi^1_0(x_1,x_2,\cdots,x_N)&=&
\frac{1}{\sqrt{N!L^N}} \mbox{det}\left[e^{2 \pi i l x_k/L }\right]_{l,k},
\nonumber\\ 
&& l=-\frac{N-1}{2},\cdots,\frac{N-1}{2}, \; k=1,\cdots, N\,,
\eea
and 
\be
A(x_j-x_i)=\cases{ 
e^{i\pi (1-\k)}& $x_j<x_i$, \cr 
1                  & $x_j>x_i$.
}  
\label{1dbraiding}
\ee

\begin{figure}[b]
\includegraphics[width=6cm]{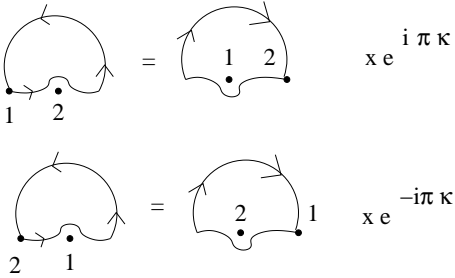}
\caption{Graphical representation of the exchange convention used in this paper
  corresponding to Eq. (\ref{1dbraiding}), or 
equivalently to Eq. (\ref{ancomm}).}
\label{braiding}
\end{figure}

The above Anyon-Fermi mapping directly generalizes the Bose-Fermi  
one \cite{g-60}.
Here, the symmetry properties of the wave function are encoded in the 
factor $\prod_{1\leq i < j \leq N} A(x_i,x_j)$ which gives the statistical 
phase $e^{i \pi(1-\k)P}$ resulting from the $P$ exchanges needed for the 
particle positions
to be brought to the ordering  $0\leq x_1< x_2< \cdots <x_N < L$.   
Note that in the definition of Eq. (\ref{1dbraiding}), we explicitly specified
the sign for the exchanging phase. 
This  amounts to fix how two anyons exchange their positions on the ring.  
Moreover, in order to define the boundary conditions, we also fixed how
a loop of one variable encircles the others. 
Figure (\ref{braiding}) shows pictorially the example of two particles once 
the convention in (\ref{1dbraiding}) is fixed.
Clearly Eq. (\ref{1dbraiding}) is equivalent to Eq. (\ref{ancomm}).

The boundary conditions on the wave function of $N$ anyons at positions 
$0 \leq x_1< x_2<\cdots x_N< L$ can then be chosen such that (for a detailed 
discussion see \cite{pka-07})  
\begin{equation}
\fl\Psi^\k_0(x_1,\cdots,x_j+L,\cdots,x_N)=e^{i \pi (1-\k) (N-1)-2i \pi (1-\k) (j-1)}\Psi^\k_{0}(x_1, \cdots,x_j,\cdots, x_N). 
\label{groundstatemono}
\end{equation}
A last remark is that, in general, one can allow an overall phase, coming for 
example from a non-zero magnetic flux penetrating the ring \cite{zw-92}.

\subsection{The one-particle density matrix as a Toeplitz determinant}

The one-particle reduced density matrix $\rho^\k_N(x_1, x_1')$ 
can be written in terms of the ground-state wave function as
\begin{equation}
\fl  \rho^\k_N(x_1,x_1')=  
\int_0^L dx_2 \int_0^L dx_3.. \int_0^L dx_N 
\overline{\Psi}^\k_{0}(x_1,x_2,x_3..,x_N)\Psi^\k_{0}(x_1',x_2, x_3,..,x_N)\,.
\label{def_oneparticle}
\end{equation}
In the case of a homogeneous system $\rho^\k_N(x_1,x_1')=\rho^\k_N(x_1-x_1')$. 
We can thus set $x_1'=0$ and study the function $\rho^\k_N(x)$ with 
$\rho^\k_N(0)=1$ as fixed by  the normalization chosen in the 
definition (\ref{def_oneparticle}), since the wave-function (\ref{AF}) 
is normalized to 1. 
From Eq. (\ref{groundstatemono}), it follows 
\begin{equation}
\rho^\k_{N}(x+L)=e^{i \pi(1- \k) (N-1)} \rho^\k_{N}(x)\,,
\label{rhoprop}
\end{equation}
i.e. $\rho^\k_N(x)$ is an $L$-periodic function only if $\k=2m/(N-1)$ with $m$
integer, as stressed in Ref. \cite{ssc-07}. 

Analogously to the case of impenetrable bosons, the anyonic one-particle 
density matrix can be written as a Toeplitz determinant.
Using  Eq. (\ref{AF}), the anyonic wave function $\Psi^\k_0$ is
\begin{equation}
\fl  \Psi_0^\k(x_1,x_2,\cdots,x_N)=\frac{1}{\sqrt{N!L^N}}\prod_{1\leq i < j \leq N} 2 i A(x_j-x_i)  \sin[\pi (x_j-x_i)/L],
\end{equation}
that allows to write the integral in Eq. (\ref{def_oneparticle}) as (in 
the angular variables $2 \pi x_j/L=t_j$)
\bea
\fl  \rho^\k_N(t)=\frac{1}{N!}\int_0^{2 \pi}d t_2\cdots\int_0^{2 \pi}d t_N
\nonumber\\ 
\prod_{s=2}^N\frac{2}{\pi} A(t_s-t)\sin[(t_s-t)/2]\sin[t_s/2]
\prod_{2\leq i<j \leq N} 4\sin[(t_j-t_i)/2]^2.
\label{onepart_int}
\eea
Note that  the dependence on the anyonic parameter $\k$ enters only
through the function $A(t_s-t)$.
Using the identity
\begin{equation}
\prod_{2\leq j<k \leq N} 4\sin[(t_j-t_k)/2]^2=\prod_{2\leq j<k \leq N} |e^{i t_k}-e^{i t_j}|^2,
\end{equation}
we can identify the second product in the integral (\ref{onepart_int})  
with the square of the absolute value of a Vandermonde determinant.
The anyonic density matrix $\rho^\k_{N}(t)=\rho^\k_{N}(2\pi x/L)$ 
can be finally written as
\begin{equation}
\rho^\k_N(t)=\frac{1}{N}\mbox{det}_{N-1}\left[\Phi^\k_{k,l}\right]\,,
\label{DET}
\end{equation}
where the $(N-1)\times (N-1)$ matrix $\Phi^\k_{k,l}$ has entries
\bea
\Phi^\k_{k,l}&=&
\frac1{2 \pi}\int_0^{2 \pi}d t_s e^{i(k-l)t_s} \phi^\k(t_s)\nonumber\\
&=&\frac{1}{2 \pi}\int_{0}^{2 \pi}d t_s e^{i(k-l)t_s} 
4 A(t_s-t)\sin[(t_s-t)/2]\sin[t_s/2]\,.
\label{final_form}
\eea
Note that we have
\bea
\Phi^1_{k,l}&=&
\frac{1}{2 \pi}\int_0^{2\pi}d t_s e^{i(k-l)t_s} 4\sin[(t_s-t)/2]\sin[t_s/2]\,,
\label{final_fermi}\\
\Phi^0_{k,l}&=&
\frac{1}{2 \pi}\int_0^{2\pi}d t_s e^{i(k-l)t_s}4|\sin[(t_s-t)/2]\sin[t_s/2]|\,,
\label{final_bos}
\eea
for $\k= 1$ and $0$ corresponding to the well known results for free fermions 
and impenetrable bosons respectively.

\subsection{The Fisher-Hartwig conjecture and the asymptotic behavior for
  large $x$}

Toeplitz matrices are fundamental objects in the study of lattice models and 
their extensive study started in the sixties for the calculation of the 
correlation functions in the classical two-dimensional Ising model. 
This study culminated with the Fisher-Hartwig conjecture \cite{fh} that 
relates the determinant of a Toeplitz matrix for large $N$ to
the analytic structure of the generating function, as briefly reviewed in the 
following.

Let us consider the $N \times N$ Toeplitz matrix ${\cal T}$ with entries
\begin{equation}
{\cal T}_{p,q}=\frac{1}{2\pi}\int_{0}^{2\pi} \phi(s) e^{i (p-q) s} d s.
\end{equation}
$\phi(s)$ is called the generating function of the matrix. 
The Fisher-Hartwig conjecture is formulated in terms of the canonical 
factorization of the generating function
\begin{equation}
\phi(s)=b(s)\prod_{r=1}^{R}T_{\beta_{r}}(s-s_r)u_{\alpha_{r}}(s-s_r)\,,
\label{factorization}
\end{equation}
where
\be
T_{\beta}(s-s_r)=\cases{
e^{-i \beta (\pi-s+s_r)}                   & $s_r< s <2\pi+s_r$, \cr
e^{-i 2\pi \beta}e^{-i \beta (\pi-s+s_r)}  & $0< s < s_r$,
}
\ee
and
\begin{equation}
u_{\alpha}(s)=(2-2\cos s)^{\alpha}.
\end{equation}
$T_{\beta}(s)$  takes in consideration the jump discontinuities while 
$u_{\alpha}(s-s_r)$ encodes the possible singularities. 
The Fisher-Hartwig conjecture states that the leading term in the limit $N\gg1$
of the determinant of ${\cal T}$ is given by
\begin{equation}
\mbox{det}\,{\cal T} \simeq 
G[b(s)]^N N^{\sum_{r=1}^{R}(\alpha_{r}^2-\beta_{r}^2)} E,
\end{equation}
where 
\begin{equation}
G[b(s)]=e^{\frac{1}{2\pi}\int_{0}^{2 \pi}d s \ln( b(s))}\,.
\end{equation}
The constant $E$ has been determined only long after the formulation of the
conjecture \cite{bm-94} 
\begin{equation}
\fl  E=\prod_{1\leq r \neq l \leq
  R}(1-e^{i(s_r-s_l)})^{-(\alpha_{r}+\beta_{r})(\alpha_{l}-\beta_{l})}\prod_{r=1}^{R} 
\frac{G(1+\alpha_{r}+\beta_{r})G(1+\alpha_{r}-\beta_{r})}{G(1+2 \alpha_{r})}\,,
\end{equation}
where $G(x)$ is the Barnes function
\begin{equation}
\fl G(z+1)=(2 \pi)^{z/2} e^{-(z+(\gamma_E+1)z^2)/2}
\prod_{k=1}^{\infty}(1+z/k)^k e^{-z+z^2/(2k)},
\end{equation} 
$\gamma_E$ is the Euler constant, $G(1)=G(2)=1$, and $G(3/2) = 1.06922\dots$.

The above result can be applied to the calculation of the one-particle density
matrix of the anyonic gas, where the generating function is given by
$\phi^\k(t_s)=4 A(t_s-t)\sin[(t_s-t)/2]\sin[s/2]$ in Eq. (\ref{final_form}).
$\phi^\k(t_s)$ is a piecewise continuous function
which takes the values
\be
\phi^\k(t_s)=\cases{
e^{i \pi (1-\k)}\sin[(t_s-t)/2]\sin[t_s/2] & for  $0<t_s<t$\,,\cr
\sin[(t_s-t)/2]\sin[t_s/2]                 & for  $t<t_s<2\pi$\,.
}
\ee
Using
\be
T_{\beta}(t_s)T_{-\beta}(t_s-t)=\cases{
e^{i \beta t+i 2 \pi \beta} &for  $t_s<t$\,,\cr
e^{i \beta t}               &for  $t_s>t$\,,
}
\ee
and
\begin{equation}
2-2\cos t_s=4\sin^2(t_s/2),
\end{equation}
the generating function can be written as
\begin{equation}
\fl\phi^\k(t_s)=e^{-i \beta t}T_{-\k/2}(t_s)T_{\k/2}(t_s-t)
(2-2\cos{t_s})^{1/2}(2-2\cos(t_s-t))^{1/2}\,.
\label{anyfactorization}
\end{equation}

Comparing Eq. (\ref{anyfactorization}) with Eq. (\ref{factorization}), 
we have $R=2$, $\beta_{1}=-\beta_{2}=-\k/2$, $\alpha_{1}=\alpha_{2}=1/2$.
Using the Fisher-Hartwig conjecture we finally find
\bea
\rho^\k_N(t)&\simeq& e^{i(N-1)\k t} N^{-1/2-\k^2/2}
[G(3/2+\k/2) G(3/2-\k/2)]^2 \nonumber\\
&&\times (1-e^{-it})^{-1/4-\k^2/4-\k/2} (1-e^{it})^{-1/4-\k^2/4+\k/2}.
\label{cool}
\eea
After simple algebraic manipulations we arrive to the final result
\begin{eqnarray}
\rho^\k_N(x)\sim && (2 N)^{-1/2-\k^2/2} [G(3/2+\k/2) G(3/2-\k/2)]^2
 \nonumber \\
   &&\times 
 e^{i \pi \k(N  x/L -1/2)}\left|\sin\left(\frac{\pi
x}{L}\right)\right|^{-1/2-\k^2/2},
\label{as_ro_any}
\end{eqnarray}
where we have reintroduced the variable $x$ via $t= 2\pi x/L$.

 For $\k=0$, Eq. (\ref{as_ro_any}) reproduces the famous Lenard result
 \cite{l2} for impenetrable bosons. For $\k=1$ instead it does {\it not} 
reduce to 
the free fermion result $\rho^1_N(x)= \sin(N\pi x/L)/(N\sin(\pi x/L))$. 
In fact, as shown, in Ref. \cite{cm-07}, Eq. (\ref{as_ro_any}) can only get the
mode at $k=-k_F= -(N - 1)/2$, but not the other one at $k = k_F$, 
that only for $\k=1$ is degenerate with the other. This is not inconsistent, 
because for $\k=1$ the Fisher-Hartwig conjecture does not hold.

All the results up to this point correspond mainly to an extended and detailed
version of those already appeared in the letter \cite{ssc-07}. 
All new results are reported in the following.

\section{Explicit calculation of the reduced density matrix and momentum
distribution function}
\label{sec3}

\begin{figure}[t]
\includegraphics[width=\textwidth]{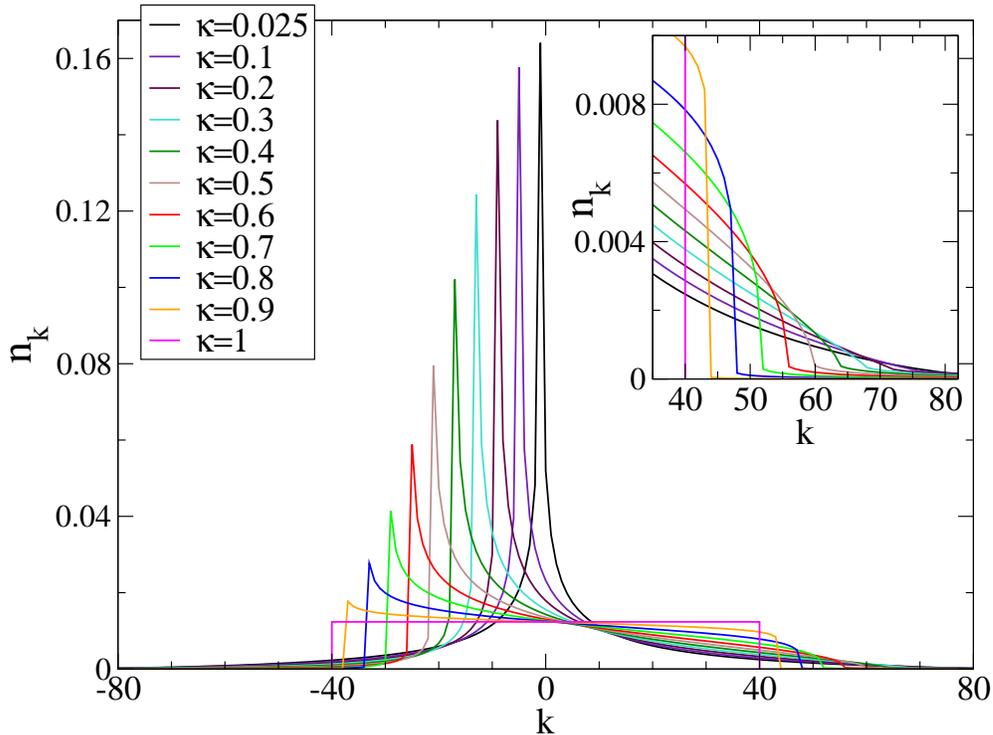}
\caption{Momentum distribution function $n^\k_{81}(k)$  (i.e. $N = L = 81$) for
several statistical parameters $\k$ showing how the peak decreases with 
increasing $\k$ and becomes the Fermi discontinuity at $k = -k_F$. 
Inset: Zoom in the region $k_F < k < 2k_F$ 
showing that the weak singularity at $k = 2k_F$ for $\k = 0$ becomes
the discontinuity of the function at $k = k_F$ for $\k = 1$.
}
\label{fig2}
\end{figure}

Eq. (\ref{DET}) provides a representation of $\rho_N^\k(x)$ 
that can be easily used to derive the correlation function at finite $N$. In
particular, for a small number of particles $N$ we have explicit expressions
of $\rho_N^\k(x)$. For $N=3$, 
using the variable $t=2 \pi x/L$, $0\leq t\leq 2\pi$, we find 
\bea
\fl \rho^\k_{N=3}(t)=
\frac{1}{24 \pi^2}\left[8\pi^2-\xi_\k(-15\xi_\k+ 8\pi t- 2\xi_\k t^2)
+\xi_\k^2 \cos 2 t \right.\nonumber\\ \left.
+4(2\pi -\xi_\k(t+2))(2 \pi -\xi_\k(t-2))\cos t+12 \xi_\k (2\pi - \xi_\k t) \sin t \right]\,,
\label{ro3}
\eea
where $\xi_\k$ is defined by
\begin{equation}
\xi_\k=1-e^{i (1-\k)\pi}\,.
\label{def_zeta}
\end{equation}
In the same manner and with the help of Mathematica,
we can expand the determinant up to $N = 11$.
However, the formulas for general $\k$ are too long to be reported here. 
Larger values of $N$ can be easily worked out numerically. The small $N$
analytic expressions are then practical cases to check numerical calculations, 
asymptotic expansions etc. as we will extensively do in
the following. For example, by fixing $t = 2\pi$ in Eq. (\ref{ro3}),  
and in the analogous ones not reported here, we can explicitely verify that
$\rho^\k_{N=3}(L)=(1-\xi_\k)^2\rho^\k_{N=3}(0)$ and 
$\rho^\k_{N=5}(L)=(1-\xi_\k)^4\rho^\k_{N=5}(0)$ etc. 
in agreement with Eq. (\ref{rhoprop}).

We now focus on the properties that can be derived from the numerical
calculations of $\rho^\k_N(x)$. The correlation function in the real space has
been already worked out numerically for several $\k$ in Ref. \cite{ssc-07}. 
The agreement with the Fisher-Hartwig result Eq. (\ref{as_ro_any}) 
has been shown to be excellent for small $\k$. When $\k$  get
closer to $1$, by using an harmonic fluid approach, it has been shown
\cite{cm-07} that a second
mode not captured by the Fisher-Hartwig, becomes important and is responsible 
for the right behavior at the fermionic point (these results were later 
rederived with a different approach \cite{pka-07}). 
Overall we can safely state that the main qualitative features 
of $\rho^\k_N(x)$ that can be extracted by numerics are well understood (with
the important exception of fixing the amplitude of the second mode, see the
discussion section for details). The same is not true for the momentum
distribution function $n^\k_N(k)$ which has only been considered marginally
\cite{ssc-07,cm-07}.

\begin{figure}[t]
\includegraphics[width=\textwidth]{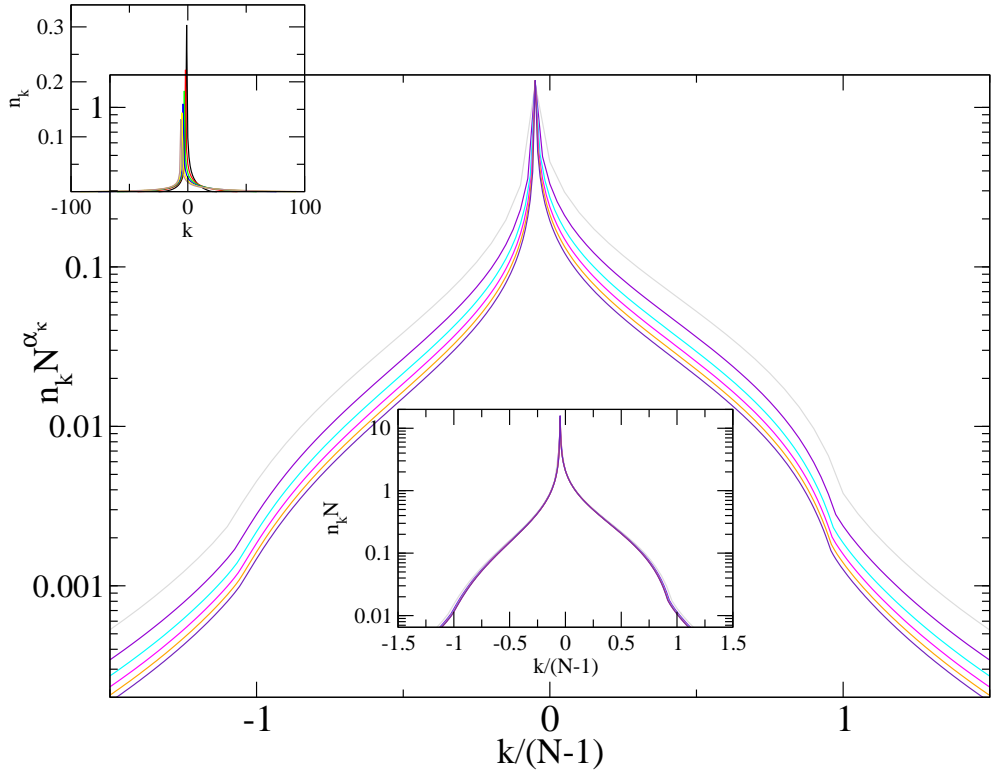}
\caption{Scaling of the momentum distribution function $n^{\k=0.1}_N(k)$ 
with varying
  $N= 21,41,61,81,101,121$. Upper inset: $n^\k_N(k)$ as function of $k$. Main
  Plot: Scaling of the peak by plotting $n^\k_N(k) N^{\a_\k}$ with
  $\a_\k=1/2+\k^2/2$ versus $k/(N-1)$. 
Central inset: $n^\k_N(k) N$ versus $k/(N-1)$ that gives the collapse 
for large $k$, and works well also for moderate values of $k$.
}
\label{fig3}
\end{figure}

\begin{figure}[t]
\includegraphics[width=\textwidth]{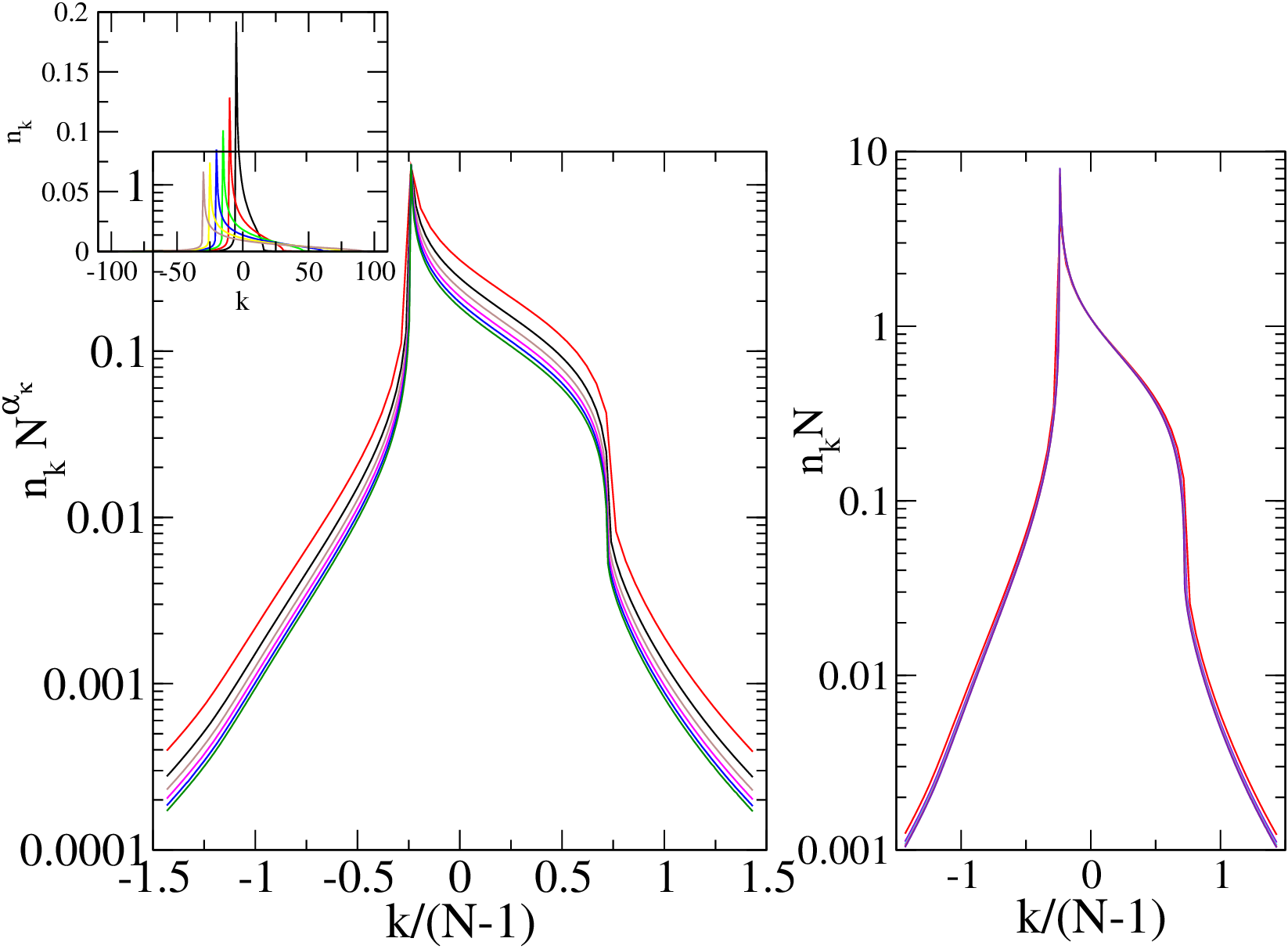}
\caption{Scaling of the momentum distribution function $n^{\k=0.5}_N(k)$ 
with varying
  $N= 21,41,61,81,101,121$. Upper inset: $n^\k_N(k)$ as function of $k$. 
Left Plot: Scaling of the peak by plotting $n^\k_N(k) N^{\a_\k}$ with
  $\a_\k=1/2+\k^2/2$ versus $k/(N-1)$. 
Right Plot: $n^\k_N(k) N$ versus $k/(N-1)$ that gives the collapse 
for large $k$.
}
\label{fig4}
\end{figure}

\begin{figure}[t]
\includegraphics[width=\textwidth]{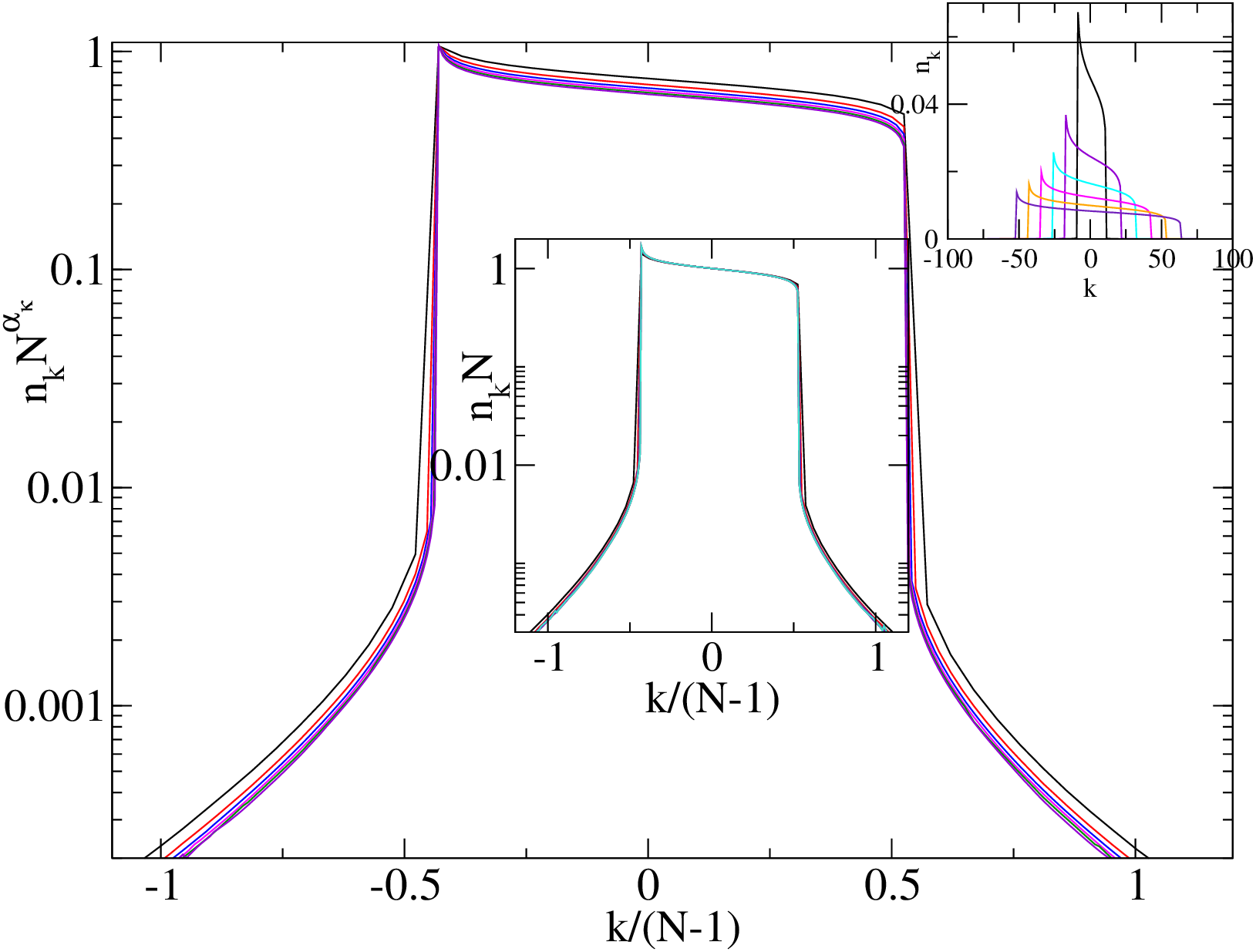}
\caption{Scaling of the momentum distribution function $n^{\k=0.9}_N(k)$ 
with varying  $N= 21,41,61,81,101,121$. 
Upper inset: $n^\k_N(k)$ as function of $k$. 
Main Plot: Scaling of the peak by plotting $n^\k_N(k) N^{\a_\k}$ with
  $\a_\k=1/2+\k^2/2$ versus $k/(N-1)$. 
Central inset: $n^\k_N(k) N$ versus $k/(N-1)$ that gives the collapse 
for large $k$.
}
\label{fig5}
\end{figure}

Here we fill this gap. For practical purposes we only consider the case of a
periodic $\rho_N^\k(x)$, i.e. $\k=2m/(N-1)$, so that $n^\k_N(k)$ 
can be defined as the Fourier transform
\be
n^\k_N(k)=\frac{1}{L}\int_0^L dx e^{2\pi i k x/L} \rho_N^\k(x)\,,
\label{fou1}
\ee
with $k$ {\it integer}. We calculated $n^\k_N(k)$ for all the values of
$\k\in[0,1]$ that give an $L$-periodic $\rho_N^\k(x)$ for $N = L =
21,41,61,81,101,121$  (higher values of $N$ could have been
easily obtained, but the values considered are sufficient for our aims). The
method is very simple: we calculated numerically $\rho_N^\k(x)$ for enough
equispaced $x$'s and then we considered the fast Fourier transform of these
data, obtaining very accurate results.

Let us first discuss the more natural question we can address with this method:
how with changing $\k$ we can smoothly interpolate between the impenetrable 
boson and the free fermion momentum distribution functions. We remember to the
reader that these two correlation functions are very different. The free
fermion one is obviously a Fermi-Dirac distribution
\be
n_N^1(k)=\cases{
1/N & for $|k|<k_F$\,,\cr
1 & for $|k|>k_F$\,,
}
\ee
with $k_F=(N-1)/2$. Oppositely at $\k = 0$, $n_N^0(k)$ has the characteristic
of a strongly interacting system with a peak in $0$, going like $N^{-1/2}$
\cite{l1} and a large-momentum
power-law tail \cite{mvt-02,od-03}. Both the limits are symmetric for $k\to-k$.

Fig. \ref{fig2} shows $n_{81}^\k(k)$ (we take $N=81$ as a typical large enough
value) for several $\k$ between 0 and 1 (the figure with all the possible $\k$ 
making $\rho_N^\k(x)$ periodic has too many curves to be readable). 
The peak at $k = 0$ for $\k = 0$ shifts backward at $k = -\k(N-1)/2$ 
and its height decreases from $N^{-1/2}$ to $N^{-\a_\k}$ 
with $\a_k=1/2+\k^2/2$. Both these
features are encoded in the Fisher-Hartwig result Eq. (\ref{as_ro_any}).
Approaching $\k = 1$ this peak becomes the discontinuity at $k = -k_F$. 
However, the Fisher-Hartwig result only
applies very close to this maximum value and cannot explain how the 
discontinuity at $k = k_F$ is produced, that instead is well understood from 
numerics. For bosons, it is known from the pioneering paper of Vaidya and 
Tracy \cite{vt-79} that not only $k = 0$ is singular, but there are additional 
weaker singularities at all the points $k = 2mk_F$ with $m$ integer. 
For example, at $k =\pm2k_F$ the second derivative of $n_N^0(k)$ is divergent 
in
the thermodynamic limit, but this singularity is so weak that is hardly seen 
in Fig. \ref{fig2}.
Increasing $\k$, all these singularities move backward of $-\k(N - 1)/2$. 
In particular, the first one at $2k_F$ moves at $2k_F -\k(N - 1)/2$ 
and becomes sharper. This is emphasized
by the inset of Fig. \ref{fig2}, where we zoom close to this region showing 
how the derivative of $n_N^\k(k)$ develops a larger discontinuity increasing
$\k$, that becomes a discontinuity of
the function itself for $\k = 1$ at $k = k_F$.

We find that this mechanism to connect smoothly $\k = 0$ to $\k = 1$ is very
interesting because of its simplicity. Furthermore we believe that it should 
be valid for the Lieb-Liniger model at arbitrary coupling, but unfortunately 
it is still impossible
to have the analytic structure of the singularities in the general case. 
However, this structure is compatible with the results from the harmonic fluid
approach \cite{cm-07,pka-07}, where 
the amplitudes that fix the strength of the singularities are free
parameters. 
Finally it is likely that this mechanism would be valid also for other models 
of interacting anyons.

There are other interesting features that can be extracted from our numerical
calculations. We notice that, in contrast with the cases $\k = 0,1$ the 
momentum distribution function is highly asymmetric. 
Increasing $\k$ the decay to the left of the peak becomes more rapid, 
while for $k > -\k(N -1)/2$, $n^\k_N (k)$ slowly
develops a plateaux that becomes the Fermi sea at $\k = 1$. 
However for value of $|k|$
larger than $k_F$ the momentum distribution function tends to restore the 
symmetry $k\to -k$. This is clear from Figs. \ref{fig3}, \ref{fig4}, and
\ref{fig5}  where we plot $n^\k_N (k)$ for $\k=0.1,0.5,0.9$ in logarithmic
scale to magnify the small values the functions take. In 
the following sections we will show rigorously that these tails follow a
$k^{-4}$ power-law and are symmetric with respect to $k = 0$ (the first 
asymmetric term is only at level $k^{-9}$ when $\k = 2m/(N -1)$).

The three Figs. \ref{fig3}, \ref{fig4}, and \ref{fig5} 
show also other interesting features of the momentum distribution function. We
plot $n^\k_N (k) N^{\a_\k}$ with $\a_\k=1/2+\k^2/2$ as function of 
$k/(N - 1) = 2k/k_F$ that should show a perfect data collapse for all the
values of $k$ well described by the Fourier transform of the asymptotic
Fisher-Hartwig result 
valid for large $x$, and so close to the peak. 
It is evident from the figures, that the
collapse is effective only extremely close to the peak. Oppositely for large
$k$,  the momentum distribution function is not expected to show any anomalous
scaling and to be proportional to $1/N$ (this will be shown rigorously for
large $k$ in the following sectionsL note the different normalizations of
$n^\k_N(x)$). For this reason we also plot $N n^\k_N(k) $ 
that in fact shows a good collapse of the data for large $k$. 
But not only: also for intermediate values of $k$ the various curves fall on 
the same ``scaling'' function.

\section{Differential equation for $\rho^\k_{N}(x)$}
\label{sec4}

An   effective  way  of  characterizing    correlation   functions   for  1D
strongly   interacting 
systems   is to  find a  differential  equation   that  the  correlation
satisfies and   from  this 
extract  the  analytical   properties   of  the solution.   For   the 1D
impenetrable    Bose   gas 
 this program    started   with  the  work   of Jimbo    et al.
 \cite{jmms-80} proving   that $\rho_{\infty}^{1}(x)$ (i.e. in the
 thermodynamic limit)   satisfies a second  order   Painlev\'e  differential
 equation   of the 
V  kind  $P_{V}$. This  result  allowed   to  derive  several  terms   in the  asymptotic    expansion
 for large  distances \cite{jmms-80}. Later  Forrester  and  collaborators
 were  able to  show  that  also the  finite $N$   correlation   function
 $\rho^0_{N}(x)$ satisfies  a  $P_{VI}$     Painlev\'e  equation   and
 they pointed   out  useful  connections    with   the  theory  of  random
 matrices.  Exploiting   this 
connection,    we show   here   that  also  the anyonic    one-particle
density  matrix $\rho^\k_{N}(x)$  can  be  characterized   via  a second
order  non-linear   differential  equation $P_{VI}$. 

The starting point is that  $|\Psi^{1}_0(x_1,x_2,\cdots x_N)|^2$ can be viewed
as a probability distribution function for some class of random  matrices
\cite{s-92}  ($\Psi^{1}_0(x_1,x_2,\cdots x_N)$ 
is the free fermion ground state). For periodic boundary condition, the
appropriate matrix ensemble is  the so called unitary circular 
ensemble \cite{metha}, for which
\begin{equation}
\fl |\Psi_0^1(x_1,x_2,\cdots,x_N)|^2= 
\frac{1}{N! L^N}\prod_{1\leq j<k \leq N} 4\sin[\pi (x_j-x_k)/L]^2=
\mbox{Ev(U(N))},
\label{pdfunitary}
\end{equation} 
 where $\mbox{Ev(U(N))}$ is the eigenvalue probability distribution function
 of unitary matrix $U(N)$ with uniform measure.   Eq. (\ref{onepart_int}) for
 $\rho^\k_{N}(x)$   can thus be interpreted as an  average in the circular
 unitary ensemble
\begin{equation}
\rho^\k_{N}(t)= \langle\prod_{s=2}^{N}\frac{2}{\pi}
A(t_s-t)\sin[(t_s-t)/2]\sin[t_s/2]\rangle_{\rm Ev(U(N))}.
\label{rho_av}
\end{equation}
It is known (see for instance  Ref. \cite{fw-02}  and references therein)
that the averages in various random matrix ensembles are related to the
Painlev\'e differential equations which are classified in six types:
$P_I,\cdots P_{VI}$.   In the case of the average  (\ref{rho_av})   it has
been shown  by Forrester et al.  \cite{ffgw-02}  that one can evaluate
$\rho^\k_{N}$  via non-linear differential equation of $P_{VI}$ type. In fact,
$\rho^\k_{2}(x),\rho^\k_{3}(x),\cdots \rho^\k_{N+1}(x)$ can be considered as a
sequence of functions $\tau_3[1](u), \tau_3[2](u),\cdots \tau_3[N](u)$ (we
introduced $\ln u= 2i x$)  which represents one  of the so called
$\tau$-function sequence  occurring  in the  $P_{VI}$ systems. The first
function of this sequence, $\tau_3[1](u)$, turns out to satisfy  a Gauss
hypergeometric  equation, which admits two independent solutions. The
parameters appearing in the hypergeometric equation do not depend on the
anyonic parameter $\k$. The effect  of the statistics is to select  a
particular solution in the bidimensional space of solutions of the
hypergeometric equation. In other words, the anyonic statistics enters only  in
the determination of the boundary conditions. Using the so called Backlund
transformations, which leave the form of the  $P_{VI}$ equations 
unchanged,   one can systematically  construct all
the $\tau_3[N]$ function from the $\tau_3[1]$ and verify that the form of the
corresponding differential equations do not depend on the
statistics. Specifically,  the  function $\sigma^\k_{N}(u)$ (directly related
to $\tau_{3}[N](u)$)   
\begin{equation}
\sigma^\k_{N}(u)=u (u-1) \ln \rho^\k_N(x(u)), \qquad x(u)=\frac{\ln u}{2 i}, 
\end{equation}
satisfies the following second order differential equation
\begin{eqnarray}
\fl u^2(u-1)^2 \left[\frac{d^2}{d u^2}\sigma^\k_N(u)\right]^2+ 
4\left[\sigma_N(u)-(u-1) \frac{d}{d u}\sigma^\k_N(u)+1\right] 
\times \label{diffeq} \\ 
\fl\times
\left[\sigma^\k_N(u)\frac{d}{du}\sigma^\k_N(u)
-u\left(\frac{d}{du}\sigma^\k_N(u)\right)^2
-\frac{N^2-1}4\left(\sigma^\k_N(u)-(u-1)\frac{d}{du}\sigma^\k_N(u)\right)\right]=0,
\nonumber
\end{eqnarray}
that in fact  does not depend explicitely on the anyonic parameter $\k$.
The study of the boundary  conditions  for  general
 values of $\k$ is missing in  Ref. \cite{ffgw-02}, where only the bosonic and
 fermionic cases ($\k=0,1$) have been discussed.
 The general $\k$ dependence of the boundary conditions ($\xi_\k$ is defined
 in Eq. (\ref{def_zeta})) 
\begin{equation}
\fl \lim_{u\to1}\sigma_{N}(u) \sim \frac{N^2-1}{12}(u-1)^2-\left(
  \frac{N^2-1}{24}+i\xi_\k\frac{N(N^2-1)}{48\pi}\right)(u-1)^3\,,
\label{bc}
\end{equation}
are derived by expanding $\rho^\k_N(t=2\pi x/L)$ in power of $x$. Let us show
how to derive this result. Using Eqs. (\ref{1dbraiding}) and (\ref{def_zeta}),
Eq. (\ref{onepart_int}) can be written as
\bea
\fl \rho^\k_{N}(t)=\frac{1}{N!}\prod_{s=2}^{N}
\left(\int_{0}^{2 \pi}d t_s -\xi_\k \int_{0}^{t}d t_s\right) 
\prod_{s=2}^{N}\frac{2}{\pi} \sin[(t_s-t)/2]\sin[t_s/2]\nonumber\\
\qquad\qquad \qquad\qquad\qquad\times
\prod_{2\leq i<j \leq N} 4\sin[(t_j-t_i)/2]^2.
\label{onepart_int2}
\eea
Expanding the above expression in powers of $\xi_\k$, $\rho_N^\k(t)$ 
is expressed
in terms of the $(n+1)$-particle density matrices  $\rho^{1}_{(n,N)}(t,
t_2,\cdots t_{n+1};0, t_2, \cdots t_{n+1})$ of a  system of $N$ free fermions
\bea
\fl \rho^\k_{N}(t)=\rho^{1}_N(t)+\nonumber \\ 
\fl \qquad +\sum_{n=1}^{N}\frac{(-\xi_\k)^n}{n!} \int_{0}^{t}d t_2 \int_{0}^{t}d t_3\cdots \int_{0}^{t}d t_{n+1} \rho^{1}_{(n,N)}(t, t_2,\cdots t_{n+1};0, t_2, \cdots t_{n+1})\,,
\label{Kform}
\eea
The $n-$th term of  the above expansion is proportional to the $n-$th power of $t$. Using the Wick  theorem, the  $(n+1)$-particles density matrix  $\rho^{1}_N(t, t_2,\cdots t_{n+1};0, t_2, \cdots t_{n+1})$ is expressed as a products of one-particle density matrices $\rho^{1}_{N}$.  Using
\begin{equation}
\rho^{1}_{N}(x)=\frac{\sin (N \pi x/L)}{N\sin (\pi x/L)},
\label{explicit_fermion}
\end{equation}  
the $2$ free fermions density matrix $\rho^{1}_{(2,N)}(t, t_2;0, t_2)$ is
\begin{eqnarray}
\fl \rho^{1}_{(2,N)}(x, x_2;0,
x_2)=\rho^{1}_N(x)\rho^{1}_N(0)-\rho^{1}_N(x_2)\rho^{1}_N(x-x_2)=
\nonumber \\\quad 
=\frac{\sin (N \pi x/L)}{N\sin \pi x/L}-\frac{\sin (N \pi x_2/L)}{N\sin (\pi x_2/L)}\frac{\sin (N \pi (x-x_2)/L)}{N\sin (\pi (x-x_2)/L)}\,.
\end{eqnarray}
At the first order in $\xi_\k$, we have
\begin{eqnarray}
\fl \rho_N(x)=\frac{\sin (N\pi x/L)}{N\sin (\pi x/L)}\nonumber \\ \fl\qquad
-\frac{\xi_\k}{L} \int_{0}^{x} d x_2 N\left[
  \frac{\sin (N\pi x/L)}{\sin (\pi x/L)}- \frac{\sin (N\pi x_2/L)}{\sin (\pi x_2/L)} \frac{\sin
   (N\pi (x-x_2)/L)}{\sin (\pi (x-x_2)/L)}\right] +O(\xi_\k^2),
\label{roexp}
\end{eqnarray}
which gives
\begin{equation}
\lim_{x\to 0} \rho_N^\k(x )= 
1-\frac{N^2-1}{6}(\pi x/L)^2+\xi_\k \frac{N(N^2-1)}{18
    \pi}(\pi x/L)^3,
\end{equation}
that is equivalent to  Eq. (\ref{bc}).

Note that Eq. (\ref{onepart_int2}) shows explicitely that the non real
(i.e. the imaginary parts) terms in $\rho^\k_{N}(x)$ come from $\xi_\k$
alone. This is suggestive for an artificial interpolation between bosons and
fermions that does not require complex phases. 

\section{Short distance and large momentum expansions}
\label{sec5}

Starting   from  the  work  of  Jimbo   et al. \cite{jmms-80}  differential
equations satisfied by correlation functions become one of the most powerful   
tool to obtain asymptotic expansions.
Eq. (\ref{diffeq}) joined with the boundary condition (\ref{bc}) allows to
obtain the small $x$ expansion of $\rho^\k_{N}(x)$, on the same line of
Ref. \cite{ffgw-03b} for impenetrable bosons. In fact, by substituting a small
$x$ power series for $\rho^\k_{N}(x)$ into the differential equation, we
obtain equations which define all but one of the coefficients. In particular
the resulting equation for the coefficient $x^3$ vanishes identically, and to
fix this parameter we require the boundary condition (\ref{bc}). With   the
help  of Mathematica, we found straightforward   to  obtain  the first 25
terms of    $\rho^\k_{N}(x)$, but it would require far too much space to
exhibit all these here. In the variable $t=2\pi x/L$ up to order $t^{10}$ we
find  
\begin{eqnarray}
\fl \rho^\k_N(x) = 
1-\frac{N^2-1}{2^2 6}t^2+\xi_\k \frac{N(N^2-1)}{2^318 \pi}t^3 
\label{smallx} \\
\fl \qquad+\frac{(3N^2-7)(N^2-1)}{2^4 360}t^4
-\xi_\k \frac{N (11 N^2-29)(N^2-1)}{2^5 2700 \pi}t^5 \nonumber \\
\fl \qquad  
-\frac{(3N^4-18N^2+31)(N^2-1)}{2^6 15120}t^6
-\xi_\k\frac{N(183 N^4-1210N^2+2227)(N^2-1)}{2^7 1587600 \pi}t^7 \nonumber \\
\fl \qquad  
+\frac{(N^2-1)[(15 N^6-165 N^4+717 N^2-1143)\pi^2+
56 N^2 (N^2-4)\xi_\k^2]}{2^8 5443200 \pi^2}t^8\nonumber \\
\fl \qquad  -\xi_\k \frac{N( -22863+13867 N^2-3017 N^4+253 N^6)(N^2-1)}{2^9
  142884000 \pi }t^9-\nonumber\\ 
\fl \qquad -\frac{N^2-1}{2^{10} 62868960000\pi^2} [ 
      525 (3 N^8-52 N^6+ 410 N^4 - 1636 N^2+2555)\pi^2 + 
\nonumber\\ \qquad 
     88 N^2(489 N^6- 4606 N^4+ 12761 N^2-8644)\xi_\k^2 )t^{10}. 
\nonumber
\end{eqnarray}
In the bosonic case, $\k=0$ ($\xi_\k=2$), we find the result obtained in
Ref. \cite{ffgw-03b}. In the fermionic case, $\k=1$ ($\xi_\k=0$), the above
formula gives the small $x$ expansion of $\rho^{1}_N(x)$ of
Eq. (\ref{explicit_fermion}). 
 
Note  that  the dependence  of the  previous  expansion   up  to the  order
$t^7$ is trivial, in the sense  that the  even  terms   are  the  same  as
for bosons,   while   the odd  ones  only get a global factor $\xi_\k$. While
the  latter property   is true at  all the orders  we  calculated, the former
has  a non-trivial   $\xi_\k$   dependence    that  firstly shows  up  in the
term  $t^8$   with a factor proportional   to $\xi_\k^2$. Higher  powers in
$t$ show   also  higher   powers  of   $\xi_\k^2$. 

\subsection{Large moment expansion}

Because of the periodicity properties  of $\rho^\k_{N}(x)$ in
Eq. (\ref{rhoprop}), the momentum distribution function cannot   be defined
simply  as  a Fourier  series.  Its definition  should  be changed according
to (for convenience we also introduced a factor $1/N$ compared to the
definition in Eq. (\ref{fou1}))
\begin{equation}
\rho^\k_N(x)=\frac{1}{N}\sum_{n=-\infty}^{\infty} n^\k_N(k) \exp \left[-2 \pi i (k+\delta_\k) \frac{x}{L} \right],
\label{fourier}
\end{equation}
where the shift $\delta_\k$ is defined by 
\begin{equation}
\delta_\k=\left\{1+(\k-1)\frac{N-1}{2}\right\},
\label{shift}
\end{equation}
and  $\{x\}=x-[x]$ stands for the non-integer part  of $x$. This shift in the
definition does not matter when   
$\rho^\k_N(x)$ is periodic, i.e. for $\k=2m/(N-1)$, when
$\delta_\k=0$. Inverting Eq. (\ref{fourier}), 
the momentum distribution function is 
\begin{equation}
n^\k_{N}(k)=\frac{N}L \int_0^L d x \exp \left[2 \pi i (k+\delta_\k) \frac{x}{L} \right] \rho^\k_N(x),
\end{equation}
that gives  the probability occupation of the states with momentum 
$2\pi (k+\delta_\k)/L$.
For a small number of particles it is possible to obtain close expressions of
$n^\k_N(k)$. For $N=3$, from Eq. (\ref{ro3}) we have (we stress that this
formula is valid only for $0<\k<1$) 
\begin{equation}
\fl n^\k_3(k)= \frac{4 \cos^2 \left( \pi\k/2 \right) \left(4\sin \left( \pi \k \right)p_2(k,\k)+ \pi p_7(k,\k)\right)}{ \pi ^3 \left(-4+k^2+2 k \k+\k^2\right) \left(-k+k^3-\k+3 k^2 \k+3 k \k^2+\k^3\right)^3},
 \end{equation}
 where the $p_2(k,\k)$ and $p_7(k,\k)$ are respectively polynomials of order
 $2$ and $7$ in $k$ 
 \begin{eqnarray}
\fl p_2(k,\k)=7 k^2 +14 k \k -1+7 \k^2, \nonumber \\
\fl p_7(k,\k)=
  3 k^7+21 k^6 \k+7 k^5 (9\k^2-2)+ 35 k^4 \k (3 \k^2-2)
+7 k^3 (1-20 \k^2+15 \k^4)+ \nonumber \\
\fl \;+7 k^2 \k (3-20 \k^2+9 \k^4)+k(4+21 \k^2-70 \k^4+21 \k^6)+
\k (4+7 \k^2-14 \k^4+3 \k^6).\nonumber
 \end{eqnarray}
For $\k=0$, this reduces to the bosonic distribution function 
 $n^{0}_3(k)$ \cite{ffgw-03b}
\begin{eqnarray}
n^{0}_{3}(0)&=&\frac{1}{3}+\frac{35}{2 \pi^2}, 
\quad n^{0}_{3}(\pm 1)=\frac{1}{3},
\quad n^{0}_{3}(\pm 2)=\frac{35}{36 \pi^2}, \nonumber \\
n^{0}_{3}(k)&=&\frac{2(3 k^7-14 k^5+7 k^3+4 k)}{(-4+k^2)(-k+k^3)^3 \pi^2}, \qquad |k|>2,
\end{eqnarray}
and for $\k=1$ to the Fermi distribution
\be
n^{1}_3(0)=n^{1}_3(\pm 1)=1, \qquad n^{1}_{3}(k)=0 \quad |k|>1\,.
\ee
The close expression for $n^\k_5(k)$  and $n^\k_7(k)$  are too long to be reported here. 

\subsection{Large momentum asymptotic expansions}

From the exact solutions for different values of $N$ ($N=3,5,7$) and $\k$,  we
can study the behavior of  the distribution  of large  momenta $k\gg1$ by
studying the asymptotic of the function  $n^\k_{N}(k-\delta_\k)$. For
$N=3$ we find 
\begin{eqnarray}
\fl n^\k_{3}(n-\delta_\k)= \frac{6\cos^2 (\pi \k/2)}{ \pi^2
  k^4}+\frac{14\cos^2 (\pi \k/2)}{\pi^2 k^6}+\frac{22\cos^2
  (\pi \k/2)}{\pi^2 k^8}+ \frac{56\cos^2 (\pi \k/2) \sin
  (\pi \k)}{\pi^3 k^9}\nonumber \\ 
\fl \qquad\qquad+\frac{30\cos^2 (\pi \k/2)}{\pi^2 k^{10}}+\frac{384\cos^2 (\pi \k/2) \sin ( \pi \k )}{ \pi^3 k^{11}}+ O(k^{-12}).
\label{asymptkn3}
\end{eqnarray} 

Note that, for generic values of $\k$, we have odd terms appearing in the
asymptotic expression (\ref{asymptkn3}). As another illustrative example of
the large $k$ behavior  of $n^\k_N(k)$, we give  the  large $k$ expansions for
$n^{0}_{5}(k)$, $n^{1/2}_{5}(k)$ and $n^{1/3}_{5}(k)$ 
\begin{eqnarray}
\fl n^{0}_{5}(k)&=& \frac{50}{  \pi^2 k^4}+ \frac{410}{  \pi^2 k^6}+ \frac{2570}{  \pi^2 k^8}+\frac{14234}{k^{10}}+0(k^{-12}), \nonumber \\
\fl n^{1/2}_{5}(k)&=& \frac{25}{  \pi^2 k^4}+ \frac{205}{  \pi^2 k^6}+ \frac{1285}{  \pi^2 k^8}+\frac{4900}{  \pi^3 k^9}+\frac{7117}{\pi^2 k^{10}} +\frac{150960}{\pi^3 k^{11}} +0(k^{-12}),  \nonumber \\
\fl n^{1/3}_{5}(k+1/3) &=&\frac{75}{2 \pi ^2 k^4}+\frac{615}{2 \pi ^2
  k^6}+\frac{3855}{2 \pi ^2 k^8}+\frac{3675 \sqrt{3}}{\pi^3
  k^9}+\frac{21351}{2 \pi ^2 k^{10}}
+0(k^{-11}).
\end{eqnarray}

For $N=7$ the large $k$  expansions for different values of $\k$ (giving periodic $\rho^\k_{N}(x)$) take the form
\begin{eqnarray}
\fl n^{0}_{7}(k)&=& \frac{196}{  \pi^2 k^4}+ \frac{3332}{  \pi^2 k^6}+ \frac{44604}{  \pi^2 k^8}+\frac{540820}{\pi^2 k^{10}} +0(k^{-12}), \nonumber \\
\fl n^{1/3}_{7}(k)&=& \frac{147}{  \pi^2 k^4}+ \frac{2499}{  \pi^2 k^6}+ \frac{33453}{  \pi^2 k^8}+\frac{86436\sqrt{3}}{  \pi^3 k^9}+\frac{405615}{\pi^2 k^{10}} +\frac{5 768 280 \sqrt{3}}{\pi^3 k^{11}} +0(k^{-12}), \nonumber \\
\fl n^{2/3}_{7}(k)&=& \frac{49}{  \pi^2 k^4}+ \frac{833}{  \pi^2 k^6}+ \frac{11151}{  \pi^2 k^8}+\frac{28812\sqrt{3}}{  \pi^3 k^9}+\frac{135205}{\pi^2 k^{10}} +\frac{1922760 \sqrt{3}}{\pi^3 k^{11}} +0(k^{-12}).\nonumber
\end{eqnarray}

The large $k$ expansion of   $n^\k_{N}(k)$ for general values of $\k$ and $N$ can be obtained by means of the Mellin transform. Given a function $f(x)$, the Mellin transform is defined as
\begin{equation}
\phi(s)=\int_{0}^{\infty} d x x^{s-1} f(x).
\end{equation}
whose  inverse is
\begin{equation}
f(x)=\frac{1}{2\pi i}\int_{c-i \infty}^{c+\infty} d s x^{-s} \phi(s),
\end{equation}
where the above notation implies a contour integral taken over  a vertical axis in the complex plane in the corresponding fundamental strip.

The Mellin transform establishes a direct mapping between the asymptotic
expansion of a function $f(x)$ near $x=0$ and the set of singularities of the
transform $\phi(s)$ in the complex plane.    This technique  has been already
applied  in Ref. \cite{ffgw-03b} to the bosonic case (i.e. $\k=0$).
 
 The asymptotic expansions we compute from the small $N$ exact solutions suggest the following general large $k$ expansion for  $n^\k_{N}(k-\delta_\k)$ 
\begin{equation}
n^\k_{N}(k-\delta_\k)\sim \frac{a^\k_1}{k^4}+\frac{b^\k_1}{k^5}+\frac{a^\k_2}{k^6}+\frac{b^\k_2}{k^7}+\frac{a^\k_3}{k^8}+\frac{b^\k_3}{n^9}+\cdots
\label{largen_exp}
\end{equation}
We use the Mellin transforms of the $\cos(2\pi n x)$ and $\sin(2\pi n x)$
functions 
\begin{eqnarray}
\fl \cos (2\pi k x)=\frac{1}{2\pi i}\int_{c-i \infty}^{c+i\infty} d s \Gamma(s)\cos(\pi s/2)  (2 \pi n x)^{-s} \quad 0<c<1,\nonumber \\
\fl \sin (2\pi k x)=\frac{1}{2\pi i}\int_{c-i \infty}^{c+i\infty} d s \Gamma(s)\sin(\pi s/2)  (2 \pi n x)^{-s} \quad -1<c<1.
\label{mellin_rep}
\end{eqnarray} 
Plugging Eq. (\ref{largen_exp}) and  Eq. (\ref{mellin_rep}) in
Eq. (\ref{fourier}), we have
\begin{eqnarray}
\fl \mbox{Re}\left[\rho^\k_{N}\right](x)&=&\frac{1}{L}n^\k_{N}(0)+\frac{1}{2\pi i} \int_{c-i \infty}^{c+i\infty}d s \,\Gamma(s)\cos(\pi s /2) (2\pi x)^{-s} g(s), \nonumber  \\
\fl\mbox{Im}\left[\rho^\k_{N}\right](x)&=&\frac{1}{2\pi i} \int_{c-i \infty}^{c+i\infty}d s \,\Gamma(s)\sin(\pi s /2) (2\pi x)^{-s} g(s),
\label{cont_int}
\end{eqnarray}
where 
\begin{equation}\fl 
g(s)=g_{\mbox{even}}(s)+g_{\mbox{odd}}(s)=\sum_{j=1}^{\infty}  a^\k_j  \zeta(2j+2+s) + \sum_{j=1}^{\infty}  b^\k_j \zeta(2j+3+s),
\label{g}
\end{equation}
with  $\Gamma(x)$ and $\zeta(x)$  the Gamma and Riemann function
respectively. By closing the contour of (\ref{cont_int}) on the left, the
above integral is  given by the sum of the residues of the functions
$\Gamma(s)\cos(\pi s/2) g(s)$ and $\Gamma(s)\sin(\pi s/2) g(s)$. The poles of
these functions  are all simple and located at the points $s=0,-2,
-3,-4,-5,\cdots$ for the integral with the cosine  and $s=-2,-3,-4\cdots$ for  
the  integral with the sine. 

In the integral (\ref{cont_int}) the singularities  arise  for  $s=-2 m$,
$m\geq 0$  from the $\Gamma(s)\cos(\pi s/2)$ function and from the $m-1$-th
term of the sum $g_{\mbox{odd}}(s)$ while for $s=-(2m+1)$, $m\geq 1$, from the
Riemann function appearing in $m$-th term  of the sum $g_{\mbox{even}}(s)$. 
Defining the  function $f_{c}(z)$ as 
\begin{equation}
 f_c(-2m)=\mbox{Res}[\Gamma(s)\cos(\pi s/2)g(s),s=-2 m], \qquad m\geq 1,
\end{equation}
the resulting expansion then reads ($t=2\pi x/L$)
\begin{eqnarray}
\fl\mbox{Re}[\rho^\k_{N}](x)=1-\frac{f_c(-2)}{4 N} t^2+\frac{4\pi a^\k_1}{6N}
t^3 +\frac{f_c(-4)}{16 N} t^4 -\frac{\pi a^\k_2}{120 N} t^5\\   
-\frac{f_c(-6)}{64 N} t^6 +\frac{\pi a^\k_3}{5040 N} t^7+\frac{f_c(-8)}{256 N} t^8-\frac{\pi a^\k_4}{362880 N}t^9+O(t^{10}).
\label{smallx_mellin}
\end{eqnarray}
The coefficients of the even terms in the above expansion contain then all the terms of the sums in $g(s)$. Note that the term arising from the residue at $s=0$ combines with $n^\k_{N}(0)/L$ to give $\rho^\k_{N}(0)=1$.
The first terms for example are 
\begin{eqnarray}
f_c(-2)&=&4 g(-2), \nonumber \\
f_c(-4)&=&\frac{4}{3}\left( g_{\mbox{even}}(-4)+\sum_{j=2}^{\infty} b^\k_j \zeta(2j +3+s)\right)+\frac{25}{9} b^\k_{1},
\end{eqnarray}
Comparing Eq. (\ref{smallx}) with Eq. (\ref{smallx_mellin}), we have  the  
$a^\k_j$ coefficients 
\begin{eqnarray}
\fl a^\k_{1}&=&\frac{N^2(N^2-1)}{24 \pi^2}\mbox{Re}[\xi_\k], \nonumber \\
\fl a^\k_{2}&=&\frac{N^2(N^2-1)(-29+11 N^2)}{720 \pi ^2}\mbox{Re}[\xi_\k],\nonumber \\
\fl a^\k_{3}&=&\frac{N^2(N^2-1)(2227-1210 N^2+183 N^4)}{40320 \pi ^2}\mbox{Re}[\xi_\k], \nonumber \\
\fl a^\k_{4}&=&\frac{N^2(N^2-1)(-22863+13867N^2-3017N^4+253 N^6))}{201600 \pi ^2}\mbox{Re}[\xi_\k]. 
\label{aj}
\end{eqnarray}
For $N=3,5,7$ we recover  the exact results given above.

In an analogous way the  coefficients $b_j^\k$ are derived from the sine
integral. Here the singularities for $s=-2 m$, $m\geq 2$, arise from the
Riemann function appearing in $(m-1)$-th term  of the sum $g_{\mbox{odd}}(s)$
while the singularities for $s=-(2m+1)$, $m\geq 1$, arise from the
$\Gamma(s)\sin(\pi s/2)$ function and from the $m$-th term of the sum
$g_{\mbox{even}}(s)$.  
Defining the  function $f_{s}(z)$ as  
\begin{equation}
 f_s(-2m-1)=\mbox{Res}[\Gamma(s)\sin(\pi s/2)g(s),s=-2 m-1], \qquad m\geq 1,
  \end{equation}
we find the following expansion ($t=2\pi x/L$)
\begin{eqnarray}
\fl \mbox{Im}[\rho^\k_{N}](x)=\frac{f_s(-3)}{8 N} t^3+\frac{\pi
  b^\k_1}{32 N} t^4 +\frac{f_s(-5)}{32 N} t^5 -\frac{\pi
  b^\k_{2}}{720 N} t^6 +\nonumber \\ 
+ \frac{f_s(-7)}{128 N} t^7 +\frac{\pi b^\k_3}{20160 N} t^8+\frac{f_c(-9)}{512 N} t^9-\frac{\pi b^\k_4}{907200 N}t^{10}+O(t^{11}).
\label{smallx_mellin_im}
\end{eqnarray}
Finally, comparing Eq. (\ref{smallx}) with Eq. (\ref{smallx_mellin_im}), 
we have the coefficients  $b^\k_j$ 
\begin{eqnarray}
b^\k_{1}&=&b^\k_{2}=0,\nonumber \\
b^\k_{3}&=&-\frac{56 N^3(N^2-1)(4-5 N^2+N^4)}{34560 \pi ^3}\mbox{Im}[\xi_\k^2], \nonumber \\
b^\k_{4}&=&-\frac{N^3(N^2-1)^2(8644-4117 N^2+489 N^4))}{201600 \pi ^3}\mbox{Im}[\xi_\k^2]. 
\label{bj}
\end{eqnarray}
We have shown that the first odd term appearing in the large $k$ expansion of
$n^\k_{N}(k)$ is the one at the order $k^{-9}$. Again, one can verify the
above values match with the exact results for a small number of particles
($N=5,7$). These odd terms are very important because they represent the onset
of the asymmetry of $n^\k_{N}(k)$ for large $k$. Note that the first
non-vanishing odd term is $k^{-9}$ because in the expansion (\ref{smallx}) the
first non trivial terms shows up at $t^8$.

\section{Large $k$ asymptotics for a finite interaction anyonic gas} 
\label{sec6}

In the case of the Bose gas, it is well understood \cite{od-03} that the 
$k^{-4}$ 
tail of the momentum distribution function (and the corresponding small $x$ 
behavior) are not a prerogative of the impenetrable limit, but are a signature 
of the delta-function interaction and so are general features of the 
Lieb-Liniger model, as confirmed by direct numerical calculations
\cite{cc-06}.
It is easy to generalize this result to the anyonic case and find the
same result.

In order to prove this, we need to briefly introduce the Bethe Ansatz solution
of the Lieb-Liniger gas \cite{k-99,bgo-06}. 
For finite arbitrary coupling $c$, the eigenstates
$\Psi^\k_c(x_1,\dots,x_N)$ can be written as \cite{k-99}
\be
\Psi^\k_c(x_1,\dots,x_N)=\Phi^\k(x_1,\dots,x_N) \Psi^0_\ct(x_1,\dots,x_N)\,,
\label{kuprod}
\ee
where the phase function
\be
\Phi^\k(x_1,\dots,x_N)=
e^{-i\pi N(N-1)\k/4}\prod_{q>p} e^{i\pi\k\e(x_q-x_p)/2}\,,
\ee
encodes all the statistic dependence of the wave functions (we adapted the 
anyon convention used in Ref. \cite{k-99} to ours).
In this way the problem is equivalent to find the bosonic wave-function 
$\Psi^0_\ct(x_1,\dots,x_N)$
(in fact this method is usually called Anyon-Boson mapping, oppositely to the
Anyon-Fermion mapping exploited in the rest of the work and valid only 
for $c = \infty$). 

Kundu \cite{k-99} showed that the bosonic coupling $\ct$ is  $\ct =
c/\cos(\pi \k/2)$ and so all the thermodynamic quantities only depends on this 
effective bosonic coupling. For example, limiting to the $L$-periodic cases, 
we can write the ground-state energy per particle as
\be
\frac{E_0}N\equiv \rho_0^2 e(\g,\k)=\rho_0^2({\tilde \gamma},\k=0)\,, 
\ee
where $\rho_0=N/L$ is the mean-density, $\g=c/\rho_0$, analogously 
${\tilde \g}$, and $e(\g,\k)$ is implicitly
defined above. For $\k = 0$, $e(\g,0)$ has not a close analytical expression, 
but it is well-known and tabulated \cite{LL}.

In Ref. \cite{od-03} the large $k$ behavior of the momentum distribution 
function
for the bosonic Lieb-Liniger model was simply derived by the fact that the wave
function at the point of contact of any two particles undergo a kink in the 
derivative proportional to the value of the eigenfunction itself. 
The modification in the anyonic case is a straightforward consequence of 
Eq. (\ref{kuprod}): only the imaginary part of the eigenfunction
$\Psi^\k_c(x_1,\dots,x_N)$ is discontinuous at a point of contact (and being
odd does not contribute to the leading term for large $k$, in analogy with the
well-known fermionic case), while the real part is continuous and its
derivative has the kink of the corresponding boson wave-function 
$\Psi^0_\ct(x_1,\dots,x_N)$. Consequently the coefficient
of the large momentum tail is the same as the one of the bosonic model 
at coupling $\ct$, i.e. \cite{od-03}
\bea
n^\k_{N\gg1}(k\to\infty)&=& 
\cos^2(\pi \k/2){\tilde \g}^2 \frac{d e(\tilde \g,\k=0)}{d \tilde \g}
\left(\frac{N}{2\pi k}\right)^4\nonumber\\
&=& \cos(\pi \k/2)\g^2 \frac{d e(\g,\k)}{d\g}
\left(\frac{N}{2\pi k}\right)^4\,,
\eea
where we adapt the result of Ref. \cite{od-03} to the normalization 
$\rho_N^\k(0) = 1$. Considering the large $\g$ expansion 
$e'(\g,\k)=\cos(\pi \k/2) 4\pi^2/(3\g^2)$ \cite{bgo-06} we agree with the
result of the previous section in the impenetrable limit.

\section{Discussions}
\label{sec7}

In this paper we presented a systematic study of the momentum distribution 
function and of the one-particle reduced density matrix of the anyonic 
generalization of the Lieb-Liniger model, obtaining the first full analytic 
description of the crossover from bosons to fermions for a strongly 
interacting model of anyons. Particular attention has
been devoted to the large momentum and small distance behavior that can be
analytical obtained with a proper generalization of the methods employed for 
impenetrable bosons. In the complementary regime (large distance) we obtained 
the leading term by applying the Fisher-Hartwig conjecture. 
Corrections to the leading behavior have been investigated numerically. 
In this regime, we find the evidence that the second mode of the harmonic 
fluid approach \cite{cm-07} plays a fundamental role in describing the
correct crossover to the free fermionic regime close to the point $k = k_F$. 
This calls for further analytical studies of the singularities of the 
momentum distribution function.
Several techniques could be used to tackle this problem. 
In the bosonic case, the
systematic large $x$ expansion has been mainly exploited via the solution 
of the second
order differential equation $P_V$ \cite{jmms-80,ffgw-03b}, but this required 
the knowledge of the proper
boundary condition for large $x$ that was available only from the mapping 
to the lattice XX model \cite{vt-79} whose equivalent is not yet known for 
anyons. An alternative method, that has been applied successfully to bosons
\cite{g-04}, would be to consider the replica approach. 
Work in this direction are in progress. Another effective way to obtain the
asymptotics (maybe even beyond the impenetrable limit) would be the approach of
Ref. \cite{fp-05}.

The knowledge of the anyonic correlation functions beyond the impenetrable 
limit
is instead still very limited. 
Only the power-law structure (and the corresponding
singularities) are known from the harmonic fluid approach \cite{cm-07} (or 
equivalently from conformal field theory \cite{pka-07}). 
A general approach would be to generalize the quantum
inverse scattering methods for bosons \cite{Kbook} to anyons and then to mix 
integrability and numerics to get the full correlation functions from the 
form-factors (on the line of Ref. \cite{cc-06}). 
But this is a very ambitious and long project. However, there is an interesting
regime where this is maybe not necessary: in Ref. \cite{bgo-06}, 
exploiting the correspondence
$\ct = c/\cos(\pi\k/2)$ \cite{k-99}, it has been argued that for 
$1 < \k < 2$ the corresponding bosonic
model is attractive, independently of the coupling constant $c$. 
The attractive Bose gas is known to form bound-states whose wave-functions 
are known with exponential
precision \cite{m-64}. This allows the explicit analytic calculation 
of the correlation functions \cite{cc-07}, 
that can eventually have some interpretation in anyonic language.

\section*{Acknowledgments}
We thank M. Mintchev and G. Shlyapnikov for useful discussions. RS thanks D.
Cabra and F. Stauffer for the collaboration in a first stage of this project This work
has been done in part when PC was a guest of the Institute for Theoretical Physics of
the Universiteit van Amsterdam (a stay supported by the ESF Exchange Grant 1311
of the INSTANS activity) and in part as guest at Ecole Normale Superieure whose
hospitality is kindly acknowledged. RS acknowledges support from ANR program
blan05-0099-01.

\end{document}